\documentclass[%
 reprint,
 amsmath,amssymb,
 aps,
 prb,
]{revtex4-2}
\usepackage{xcolor}
\usepackage{graphicx}% Include figure files
\usepackage{dcolumn}% Align table columns on decimal point
\usepackage{bm}% bold math
\begin{document}

%\preprint{APS/123-QED}

\title{Helicons in multi-Weyl semimetals}% Force line breaks with \\

%\author{Panchlal Prabhat}
% \altaffiliation[Also at ]{Physics Department, L.N. Mithila University}%Lines break automatically or can be forced with \\
%\author{Amit Gupta}%
%\affiliation{%
 %Department of Physics, L.N. Mithila University, Darbhanga\\


\author{Shiv Kumar Ram}
\affiliation{ Department of Physics, Lalit Narayan Mithila University, Darbhanga, Bihar 846004, India}%

\author{Amit Gupta}
\affiliation{
  Department of Physics, M. R. M. College, Lalit Narayan Mithila University, Darbhanga, Bihar 846004, India}

%\affiliation{
 %Third institution, the second for Charlie Author
%}%


\date{}% It is always \today, today,

\begin{abstract}
Helicons are low-frequency, circularly polarized transverse electromagnetic modes that propagate in three-dimensional conducting systems in the presence of a static magnetic field. Here, we investigate helicon excitations in gapless multi-Weyl semimetals (mWSMs) with topological charge \(J=1,2,\) and \(3\) using semiclassical Boltzmann transport theory. We incorporate both Berry-curvature and orbital-magnetic-moment corrections into the quasiparticle dynamics and determine the optical-conductivity tensor in the presence of a weak static magnetic field. The longitudinal conductivity is obtained analytically for single-, double and triple-Weyl systems, whereas the transverse conductivity is evaluated numerically in the low-frequency regime relevant to helicon propagation. We find that the cyclotron frequency acquires a nontrivial dependence on the topological charge and on the magnetic field. The resulting helicon dispersion retains the characteristic quadratic dependence on the wave vector, while the axion-electrodynamics contribution generates an additional linear term. We further show that the axion term lifts the degeneracy of the gapped collective modes at zero wave vector, with the resulting mode frequencies determined by the \(J\)-dependent plasma frequencies. These results demonstrate that the helicon response provides a probe of the distinct electrodynamic properties of multi-Weyl semimetals.
\end{abstract} 

\maketitle

%\tableofcontents

\section{\label{sec:level1}
Introduction
%\protect\\ The line
 }
Helicons are low-frequency, circularly polarized transverse electromagnetic waves that propagate in a three-dimensional (3D) conducting medium in the presence of a static magnetic field $\mathbf B$ \cite{kittel2018introduction, jackson2021classical, konstantinov1960possible}. In conventional metals and plasmas, their dispersion is quadratic in the wave number in the long-wavelength regime \cite{platzman1973waves}
\begin{equation}
\omega=\frac{c B k^2}{4\pi n_e e}
\end{equation}

\noindent where $n_e$ is the electron density and $k=|\mathbf k|$ is the wave-number magnitude. Helicons can also hybridize with collective charge oscillations, producing helicon--plasmon polaritons and additional features in the optical response.

Weyl semimetals (WSMs) are three-dimensional topological materials in which the conduction and valence bands touch at isolated points in momentum space, known as Weyl nodes \cite{wan2011topological, burkov2011weyl, xu2011chern, huang2015weyl, lv2015observation, xu2015discovery}. Each Weyl node acts as a monopole of Berry curvature and carries an integer topological charge. In addition to their bulk topological properties, WSMs can host surface Fermi-arc states connecting the surface projections of Weyl nodes with opposite chirality. The electromagnetic response of a WSM can be modified by a topological magnetoelectric coupling. In a low-energy description, this contribution is represented by an axion term proportional to $\theta(\mathbf r,t)\,\mathbf E\cdot\mathbf B$, where the axion angle depends on the momentum- and energy-space separation of the Weyl nodes \cite{pellegrino2015helicons}. This term can modify electromagnetic-mode propagation. Pellegrino et al. \cite{pellegrino2015helicons} studied helicons in WSMs and showed that the topological contribution changes the mode dispersion.  

\begin{equation}
\omega = \frac{k^{2}Bc}{4\pi n_{e}e}+\frac{\alpha b_0 B k}{2\pi^2 n_e e}.
\end{equation}
where $\alpha=e^2/(\hbar c)\approx 1/137$ is the fine-structure constant and $b_0$ parameterizes the energy-space separation of the Weyl nodes. The second term originates from the axion contribution. Strain-induced pseudomagnetic fields can also modify helicon-like excitations, often referred to as pseudohelicons \cite{gorbar2017pseudomagnetic}.

Multi-Weyl semimetals (mWSMs) generalize ordinary WSMs by hosting nodes with topological charge $J>1$. Their low-energy dispersions are anisotropic: the energy is linear along one momentum direction and nonlinear in the perpendicular plane \cite{xu2011chern, huang2015weyl, huang2016new}. These systems exhibit a range of unusual transport and optical phenomena \cite{ahn2017optical, mukherjee2018doping, nag2020thermoelectric, nag2020magneto, menon2020anomalous, gupta2019novel, gupta2017floquet, gupta2022kerr}. It is therefore natural to ask how their distinct band geometry and higher topological charge modify helicon excitations.\\

In this work, we investigate helicons in gapless mWSMs with topological charges $J=1,2,$ and $3$. We use a semiclassical Boltzmann approach that incorporates Berry-curvature and orbital-magnetic-moment (OMM) corrections to calculate the optical-conductivity tensor in a weak static magnetic field. We obtain analytical expressions for the longitudinal conductivity and evaluate the transverse conductivity numerically in the low-frequency regime relevant to helicon propagation. We find that the cyclotron frequency depends nontrivially on the topological charge, and the characteristic helicon dispersion retains both quadratic and axion-induced linear dependence on the wave vector. In addition, the axion contribution lifts the degeneracy of the gapped collective modes at zero wave vector. The corresponding mode frequencies are controlled by the (J)-dependent plasma frequencies $\omega_{p,J}$, providing a direct electrodynamic signature of the multi-Weyl character. The remainder of the paper is organized as follows. Section II \& III introduces the axion-modified Maxwell equations and the semiclassical Boltzmann formalism, including OMM corrections respectively, and derives the optical-conductivity tensor. Section IV discusses the helicon dispersion and the gapped collective modes. Section V summarizes the main results.

% AUTHOR CHECKS BEFORE SUBMISSION:
% (i) Verify all axion-modified Maxwell equations and the displacement-field
%     expression against one consistent Gaussian-unit convention.
% (ii) State the collision/relaxation-time assumption in the Boltzmann equation.
% (iii) Recheck the signs, factors of hbar and c, and field-corrected energy in
%       the semiclassical equations and OMM definitions.
% (iv) Re-derive the dielectric-tensor dispersion relation from the full wave
%      equation for the stated propagation geometry.
% (v) The three zero-wave-vector branches are not generally mutually degenerate
%     when omega_{+,J}, omega_{-,J}, and omega_{p,J}/sqrt(epsilon_infty) differ.
% (vi) Check dimensions and units in the quoted plasma frequency and all figures.
\section{Maxwell's equation in WSMs}
The axion angle is taken to be $\theta(\mathbf r,t)=2(\mathbf b\cdot\mathbf r-b_0t)$, where \(\mathbf b\) and \(b_0\) characterize the momentum- and energy-space separations of the Weyl nodes, respectively. This axion field introduces an additional contribution, \(\mathcal L_\theta\), to the conventional Maxwell Lagrangian \(\mathcal L\) \cite{pellegrino2015helicons, zyuzin2012topological}, such that

\begin{equation}
\mathcal{L} =\frac{1}{8\pi}(E^{2}-B^{2})-\rho \varphi+\textbf{J}\cdot\textbf{ A} + \mathcal{L}_{\theta}
\end{equation}
where
\begin{equation}
\mathcal{L}_{\theta}=-\frac{\alpha}{4\pi ^{2}} \theta(\textbf{r},t)\textbf{E}\cdot \textbf{B}
\end{equation}

The axion term $\theta(\textbf{r},t)$ modified two of the Maxwell's equations

\begin{equation}
\nabla \cdot {\bm D} = 4\pi(\rho + \frac{\alpha}{2\pi^{2}} \textbf{b }\cdot \textbf{B})
\end{equation}

\begin{equation}
\nabla \cdot {\bm B} = 0
\end{equation}

\begin{equation}
\nabla \times {\bm E}= -\frac{1}{c}\frac{\partial {\bm B}}{\partial t}
\end{equation}
and
\begin{eqnarray}
\nabla \times {\bm{ B}}-\frac{\partial \bm{D}}{\partial t}
=\frac{4\pi}{c}(\bm{J}-\frac{\alpha}{2\pi^2} \bm{b}\times \bm{E} +\frac{\alpha}{2\pi^{2}} b_{0} \bm{B})
\end{eqnarray}

Using these modified Maxwell equations, the displacement field can be expressed as

\begin{eqnarray}\label{disp1.eqn}
\bm{D}=(\epsilon_\infty +\frac{4\pi i \sigma}{\omega}) \bm{E}+\frac{ i\alpha }{\pi \omega}\dot{\theta}\bm{B}+\frac{i c\alpha}{\pi \omega} (\nabla\theta)\times \bm{E}
\end{eqnarray}

\noindent Here, \(\epsilon_\infty\) denotes the background dielectric constant and \(\sigma\) is the electrical conductivity. The first term describes the combined background dielectric response and conductivity contribution, whereas the remaining terms represent the axion-induced electromagnetic response. The magnetic field in Eq.(\ref{disp1.eqn}) is related to the electric field by Faraday's law of induction $\bm B=\frac{c}{i\omega}\nabla \times \bm E$. Accordingly, the wave equation for the electric field propagation is modified to be

\begin{eqnarray}\label{wave_equation}
\nabla \times( \nabla \times \bm{E})+\frac{1}{c^{2}}\frac{\partial^2 }{\partial t^2}\bm{D}=0
\end{eqnarray}

\section{Model Hamiltonian and Semiclassical Boltzmann approach}
Near a node with chirality $s$ and topological charge of magnitude $J$, we use the low-energy continuum Hamiltonian \cite{dantas2018magnetotransport, nandy2021chiral, medel2024electric}
%%%%%%%%%%%%%%%%%
\begin{eqnarray} 
\mathcal{H} ( \mathbf k) = \hbar
\mathbf d_s( \mathbf k) \cdot \boldsymbol{\sigma} \,,
\end{eqnarray}
with 
$ k_\perp=\sqrt{k_x^2 + k_y^2}$ and
$\phi_k=\arctan({\frac{k_y}{k_x}})$ 

\begin{eqnarray}
\mathbf d_s( \mathbf k) &=
\left \lbrace
\alpha_J \, k_\perp^J \cos(J\phi_k), \,
\alpha_J \, k_\perp^J \sin(J\phi_k), \,
s \, v_F \, k_z \right \rbrace,
\end{eqnarray}

\noindent Here, $\boldsymbol{\sigma}=(\sigma_x,\sigma_y,\sigma_z)$ denotes the vector of Pauli matrices, $s=\pm1$ labels the node chirality, $v_F$ is the velocity along the $z$ direction, and $\alpha_J$ is a material-dependent parameter controlling the in-plane dispersion. For $J=1$, choosing $\alpha_1=v_F$ recovers an isotropic Weyl cone; for $J=2$, $\alpha_2$ is related to the inverse effective mass. The eigenvalues of the Hamiltonian are given by
\begin{align} 
\varepsilon_{\bm k}
=  \chi \hbar\sqrt{\alpha_J^2 \, k_\perp^{2J} + v_F^2 \, k_z^2}\,,
\end{align}
Here, $\chi=+1$ ($-1$) labels the conduction (valence) band. The isotropic linear Weyl dispersion is recovered for $J=1$ and $\alpha_1=v_F$.

For a Weyl node of chirality \(s=\pm1\), the carrier dynamics is described by the semiclassical Boltzmann equation. This approach is valid when the characteristic frequencies satisfy \(\omega,\omega_c\ll\varepsilon/\hbar\). Throughout this work, we consider the collisionless regime, characterized by \(\omega\tau\gg1\). The Boltzmann equation is therefore written as\\

\begin{eqnarray}
\frac{\partial \tilde{f}^s}{\partial t}+\dot{\bm k}.\frac{\partial \tilde{f}^s}{\partial \bm k}+\dot{\bm r}.\frac{\partial \tilde{f}^s}{\partial \bm r}=0\label{vlasov}
\end{eqnarray}

Here, $\tilde f^s$ denotes the distribution function for carriers near a node of chirality $s$.

In the presence of a static magnetic field $\bf{B}$ and a time varying electric field $\bf{E}$, the semiclassical equations of motion are
\begin{eqnarray}
\dot{\bm r}=\frac{1}{\hbar}\nabla_{\bm k}\tilde{\varepsilon}_{\bm k}^{s}-\dot{\bm k}\times{\bm{\Omega}}_{k}^{s}\label{EoMa}\\
\hbar \dot{\bm k}=-e\bm{E}-\frac{e}{c}\dot{\bm r}\times \bm{B}\label{EoMb}
\end{eqnarray}
Here, -e is the electron charge and \(\boldsymbol{\Omega}_{\mathbf k}^{s}\) denotes the Berry curvature. The first term on the right-hand side of Eq. (\ref{EoMa}) represents the group velocity, defined in terms of the field-corrected quasiparticle energy \(\tilde\varepsilon_{\mathbf k}^{s}\). This energy includes the orbital magnetic moment (OMM) correction and is expressed as
\(
\tilde\varepsilon_{\mathbf k}^{s}
=\varepsilon_{\mathbf k}^{s}+\varepsilon_{\mathbf k}^{m,s},
\qquad
\varepsilon_{\mathbf k}^{m,s}
=-\mathbf m_{\mathbf k}^{s}\cdot\mathbf B.
\)
Thus, the OMM modifies the quasiparticle energy and, consequently, the electronic response \cite{pellegrino2015helicons}. The orbital magnetic moment originates from the self-rotation of a Bloch wave packet and produces a magnetic-field-dependent shift in its energy \cite{sundaram1999wave}. Accordingly, the group velocity can be decomposed into its conventional and OMM-induced contributions:
\begin{eqnarray}
\mathbf v_{\mathbf k}^{s}
\equiv\frac{1}{\hbar}\nabla_{\mathbf k}\tilde\varepsilon_{\mathbf k}^{s}
=\mathbf v_s^{(0)}(\mathbf k)+\mathbf v_s^{(m)}(\mathbf k),
\end{eqnarray}
where
\(
\mathbf v_s^{(0)}(\mathbf k)
=\frac{1}{\hbar}\nabla_{\mathbf k}\varepsilon_{\mathbf k}^{s},
\qquad
\mathbf v_s^{(m)}(\mathbf k)
=\frac{1}{\hbar}\nabla_{\mathbf k}\varepsilon_{\mathbf k}^{m,s}.
\)

The Berry curvature $\bm{\Omega}_{\bm k}^s$ and the orbital magnetic moment $\bm{m}_{\bm k}^s$ are defined as \cite{sundaram1999wave, xiao2010berry, gao2022suppression}

\begin{eqnarray}
\bm{\Omega}_{\bm{k}}^s&=&Im[\langle \nabla_{\bm k} u_k^s\vert \times \vert \nabla_{\bm k} u_k^s\rangle]\\
\bm{m}_{\bm{k}}^s&=&-\frac{e}{2\hbar c}Im[\langle \nabla_{\bm k} u_k^s\vert \times (\mathcal{H}_J(\bm k)- \varepsilon_{\bm k}^s)\vert \nabla_{\bm k} u_k^s\rangle] 
\end{eqnarray}
where \(\vert u_{\mathbf k}^{s}\rangle\) is the periodic part of the Bloch wave function satisfying $ \mathcal{H}_J(\bm k)\vert u_k^s\rangle= \varepsilon_{\bm k}^s\vert u_k^s\rangle $.\\

The general expressions for Berry curvature and orbital magnetic moment for mWSMs are \cite{nandy2021chiral}
\begin{eqnarray}
\bm{\Omega}_{\bm{k}}^s =- \frac{s}{2} \frac{J v_F \alpha_J^2 k_{\perp}^{2J-2}}{\beta_{\bm k,s}^3}\{k_x,k_y,J k_z\}\\
\bm{m}_{\bm{k},s}^{\pm}=-\frac{s}{2} \frac{e J v_F \alpha_J^2 k_{\perp}^{2J-2}}{c \beta_{\bm k,s}^2}\{k_x,k_y,J k_z\}
\end{eqnarray}
where $\beta_{\bm k,s}=\sqrt{\alpha_J^2 k_\perp^{2J}+\mathit{v}_F^2 k_z^2}$. \\

%From these expressions, we immediately observe the identity
%\begin{align}
%\bm{m}_{\bm{k},s}
%= \frac{e \beta_{\bm k,s}}{c}\,\bm{\Omega}_{\bm{k}}^s \, . 
%\end{align}
%While the BC changes sign with $s$, the OMM does not.

Solving the coupled equations of motion, Eqs. (\ref{EoMa}) and (\ref{EoMb}), yields
\begin{equation}
\dot{\bm{ r}}=\frac{1}{\hbar D}\Bigl[\nabla_{\bm{ k}}\tilde{\varepsilon}_{ \bm{k}}^s + e  \bm{E} \times \bm{\Omega}_{\bm k}^s + \frac{e}{\hbar c} (\nabla_\textbf{k}\tilde{\varepsilon}_{\bm{k}}^{s}\cdot\bm{\Omega}_{\bm{k}}^{s} )\bm{B}\Bigr]
\label{eqforvel}
\end{equation}

\begin{equation}
\dot{ \bm{k}}=\frac{1}{\hbar D}[-e \bm{E}-\frac{e}{\hbar} \nabla_{\bm{ k}}\tilde{\varepsilon}_{ \bm{k}}^s\times \bm{B}-\frac{e^2}{\hbar c}(\bm{E} \cdot \bm{B})\bm{\Omega}_{\bm k}^{s}]
\end{equation}

Here,
\(D=1+\frac{e}{\hbar c}(\boldsymbol{\Omega}{\mathbf k}^{s}\cdot\mathbf B \)
is the semiclassical phase-space factor. In the weak-magnetic-field regime, its inverse can be expanded as
\noindent where $D=1+\frac{e}{\hbar c}(\mathbf\Omega^s_{\mathbf k}\cdot\mathbf B)$ is the semiclassical phase-space factor. In the weak-field regime, its inverse may be expanded as
\begin{eqnarray}
D^{-1}=1- \frac{e}{\hbar c}(\bm{\Omega}_{\bm k}^{s} \cdot \bm B) + \Bigl(\frac{e}{\hbar c}\Bigr)^2(\bm{\Omega}_{\bm k}^{s} \cdot \bm B)^2+....
\end{eqnarray}

To solve the Boltzmann equation Eq. (\ref{vlasov}), we expand the distribution function to first order in the time-dependent electric field:
\begin{equation}
\tilde{f}^{s}=\tilde{f}_{0}^{s}+\tilde{f}_{1}^{s} e^{-i\omega t} \label{distr}
\end{equation}

where $\tilde{f}_{1}^{s}$ is linear in $ \bf{E} $ and is parametrized as follows

\begin{equation}
\tilde{f}_{1}^{s}=-\frac{\partial\tilde f_{0}^{s}}{\partial \varepsilon_{k}^{s}}( X_{-}e^{i \phi}+ X_{+}e^{-i \phi} + X_{0}) \label{distr_exp}
\end{equation}

The $\tilde{f}_{0}^{s}( \tilde{\varepsilon}_{\bm k}^s)$ can be expanded at a low magnetic field as
\cite{pellegrino2015helicons}

\begin{eqnarray}
\tilde{f}_{0}^{s}(\tilde{\varepsilon}_{\bm k}^s)=\tilde{f}_{0}^{s}( \varepsilon_{\bm k}^s -\bm{m}_{\bm{k}}^s\cdot \bm{B})\nonumber\\
\simeq \tilde{f}_{0}^{s}( \varepsilon_{\bm k}^s)-\bm{m}_{\bm{k}}^s\cdot \bm{B}\frac{\partial \tilde{f}_{0}^{s}( \varepsilon_{\bm k}^s)}{\partial \varepsilon_{\bm k}^s}
\end{eqnarray}
Using Eqs. (\ref{eqforvel}) and (\ref{distr_exp}), the linear-response current density is obtained as
\begin{eqnarray}\label{cur_den}
\bm{j}_1=-\frac{e}{(2\pi)^3}\int d^3k \Bigl[\bm{\mathit{\tilde{v}}}_{\bm k}^s+\frac{e}{\hbar c}(\bm{\Omega}_{\bm k}^s \cdot \bm{\mathit{\tilde{v}}}_{\bm k}^s )\textbf{B}\Bigr]\tilde{f}_1^{s} 
\end{eqnarray}

The above equation can be expressed in frequency space $\omega$ as
\begin{equation}
j_a(\omega)=\sigma_{ab}(\omega)E_b(\omega)
\end{equation}

Substituting Eqs. (\ref{distr}) and (\ref{distr_exp}) into Eq. (\ref{vlasov}) gives

\begin{eqnarray}
\ X_{\pm}&=&\frac{e}{2D}\frac{v_{k_\perp}^s (E_{x}\pm iE_{y})}{ i[\omega\pm\frac{e Bv_{k_\perp}^s }{c D k_\perp}] }\\
&=& \frac{e}{2D}\frac{v_{k_\perp}^s (E_{x}\pm iE_{y})}{ i[\omega\pm\omega_{c_J}] }
\end{eqnarray}
and 
\begin{equation}
X_{0}=\frac{e E_{z} }{ i \omega D }[\tilde{v}_{kz}+\frac{e B}{\hbar c}( \bm{\Omega}_{\bm{k}}^s\cdot \bm{\tilde{v}}_{\bf k}^s)]
\end{equation}
\noindent where $k_{\perp}=\sqrt{k_x^2+k_y^2}$\\
and 
\begin{eqnarray}\label{cyclotron_freq}
 \omega_{c_J}=\frac{e B v_{k_\perp}^s }{ c D k_\perp} 
\end{eqnarray}
 Equation (\ref{cyclotron_freq}) gives the cyclotron frequency within the semiclassical approximation used here.\\

Exploiting the azimuthal symmetry about the (z) axis, we introduce cylindrical coordinates \cite{medel2024electric},
\begin{align}
k_{x} = k_{\perp} \cos \phi \,, \quad
k_{y} = k_{\perp} \sin \phi\ \text{ and } k_{z} = k_{z}\,, 
\end{align}
where \(k_\perp\in[0,\infty)\) and \(\phi\in[0,2\pi)\). In this coordinate system, the transverse components of the group velocity can be expressed as

\onecolumngrid
\begin{center}\
\begin{figure}
\begin{tabular}{ccc}
\includegraphics[width=0.35\textwidth]{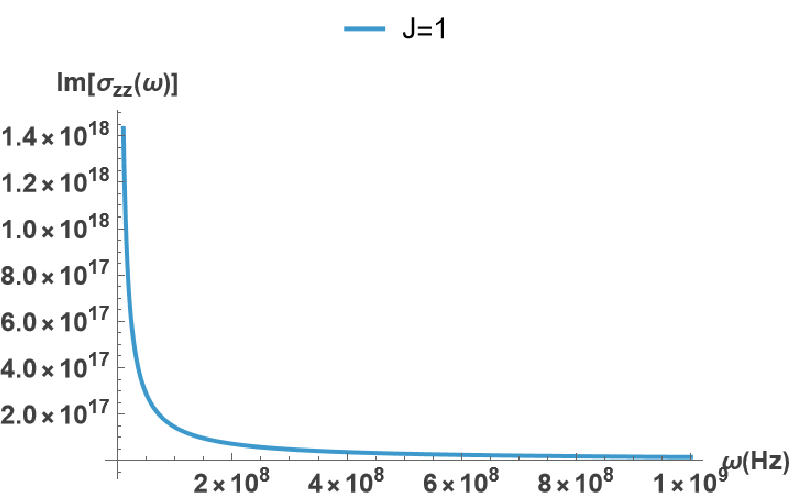} &
\includegraphics[width=0.35\textwidth]{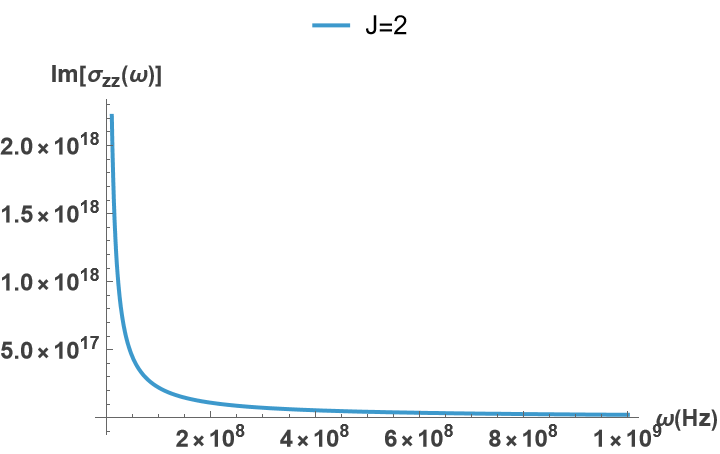} &
\includegraphics[width=0.35\textwidth]{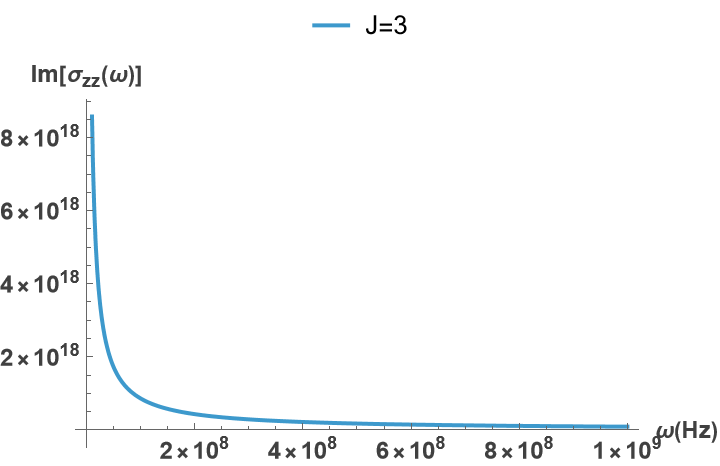} \\
\end{tabular}
\caption{The frequency dependence of the longitudinal optical
conductivity at  B = 0.01 T. The other parameters are taken as $v_F =3\times 10^7 cm/s$, $\epsilon_F$ = 5 meV, $\alpha_2 = 90 cm^2/s $ and $\alpha_3= 4.5 \times 10^{-5} cm^3/s$.} 
\label{long_cond}\
\end{figure}\
\end{center}\
\twocolumngrid

\begin{equation}
 v_{k_x}^s=v_{k_\perp}^s \cos\phi, v_{k_y}^s=v_{k_\perp}^s \sin\phi
\end{equation}
with

\begin{eqnarray}\label{velocity_plan}
v_{k_\perp}^s&=&\frac{\alpha_J ^2 J \hbar  k_{\perp}^{2 J-3}}{c \varepsilon_{\bm k}^4} \nonumber\\&&\left[c k_{\perp}^2 \varepsilon_{\bm k}^3 + s B e J k_z v_F \hbar ^2 \{(J-1) k_z^2
v_F^2-\alpha_J ^2 k_{\perp}^{2 J}\}\right]\nonumber\\
\end{eqnarray}
Combining Eqs.~(\ref{cyclotron_freq}) and (\ref{velocity_plan}), we find that, for (J>1), the cyclotron frequency depends on the transverse momentum \(k_\perp\), indicating that it is sensitive to the location of the cyclotron orbit on the Fermi surface.

In the next step, we change variables from $(k_{\perp} , k_{z})$ to $(\varepsilon_{\bf k} , \theta )$ by the coordinate transformation \cite{mandal2025anisotropic}
\begin{align}
k_{\perp} = \left( \frac{\varepsilon_{\bm k}}{\hbar \alpha_{J} } \sin \theta \right)  ^{1/J} , \qquad  k_{z} = \frac{\varepsilon_{\bm k}}{\hbar v_F} \cos \theta ,  
\end{align}
where $\varepsilon_{\bm k} \in [0, \infty )$ and $\theta \in [0, \pi )$. The Jacobian of the transformation is $\mathcal{J} (\varepsilon_{\bm k} , \theta) =   \frac{1}{J v_F \hbar \sin \theta } \left( \frac{\varepsilon_{\bm k} \sin \theta}{\hbar \alpha_{J} } \right) ^{1/J} $, leading to analytical expressions for the longitudinal conductivities $\sigma_{zz}^{(J)}$ of mWSMs

\begin{equation}\label{cond_zz_swsm}
\sigma_{zz}^1(\omega)
=i \frac{e^2 \varepsilon_F^2
}{3 \pi^2 v_F \omega \hbar^3}\left(1+ \frac{13  B^2 e^2 v_F ^2 \hbar ^2}{20 c^2 \varepsilon_F ^4}\right)
\end{equation} 

\begin{equation}\label{cond_zz_dwsm}
\sigma_{zz}^2(\omega)
=i\frac{e^2 v_F \varepsilon _F}{8 \pi  \alpha_2 \omega  \hbar ^2} \left(1+\frac{9 B^2 e^2  \alpha _2^2}{2 c^2 \varepsilon _F^2}\right)
\end{equation}
and

\begin{eqnarray}\label{cond_zz_twsm}
\sigma_{zz}^3(\omega)
=i \frac{4301 e^2 v_F \varepsilon _F^{2/3}}{2592 \pi ^{3/2}\omega  \hbar ^{5/3}\alpha_3^{2/3}} \left(1-\frac{243 B^2 e^2\alpha_3^{4/3}  \Gamma \left(\frac{5}{3}\right)\Gamma \left(\frac{29}{3}\right)}{391 c^2 \hbar^{2/3}\varepsilon_F^{4/3} \Gamma\left(\frac{1}{3}\right)\Gamma\left(\frac{25}{6}\right)}\right)\nonumber\\
\end{eqnarray}

Equation~(\ref{cond_zz_swsm}) reduces to the longitudinal conductivity obtained in Ref.~\cite{pellegrino2015helicons}. The frequency dependence of the longitudinal optical conductivity for single-, double-, and triple-Weyl semimetals is presented in Fig.~(\ref{long_cond}). The numerical results are obtained using the parameters specified in the figure caption. The analytical expressions exhibit a characteristic dependence on the Fermi energy and the topological charge. In particular, the conductivity scales with the Fermi energy according to the power law \(\varepsilon_F^{2/J}\), with additional dependence on the dispersion parameters. The variation of the longitudinal conductivity with the topological charge reflects the anisotropic energy dispersion and the associated modification of the density of states \cite{prabhat2026magneto}.\\

We now turn to the transverse components of the optical conductivity. For this purpose, we first simplify the general expression for the cyclotron frequency by retaining terms up to second order in the magnetic field. The resulting expression is

\begin{widetext}
\begin{eqnarray}\label{cyclo_freq}
 \omega_{c_J} = \frac{B e J \alpha_J ^{2/J}  \hbar ^{\frac{2}{J}-2} \sin ^{2-\frac{2}{J}}(\theta )}{c \varepsilon_F ^{\frac{2}{J}-1}}+\frac{B^2 e^2 J^2 s \cos (\theta ) \alpha_J ^{4/J}  \hbar ^{\frac{4}{J}-3} \sin^{2-\frac{4}{J}}(\theta ) \left((J-2) \varepsilon_F ^2 \sin ^2(\theta )+2 (J-1) \varepsilon_F ^2 \cos ^2(\theta )\right)}{2c^2 \varepsilon_F ^{\frac{4}{J}+1}} 
\end{eqnarray}
\end{widetext}

The first term in Eq.~(\ref{cyclo_freq}) represents the conventional semiclassical cyclotron frequency, whereas the second term describes the correction arising at quadratic order in the magnetic field. At the Fermi surface, \(\varepsilon_{\bm k}=\varepsilon_F\), the cyclotron frequency depends explicitly on the polar angle \(\theta\). Consequently, the cyclotron frequency is sensitive to the location of the orbit on the Fermi surface. Its dependence on the topological charge is therefore not determined by \(J\) alone but also involves the chemical potential, the angular coordinates, and the dispersion parameters.\\

For a single Weyl semimetal (\(J=1\)), Eq.~(\ref{cyclo_freq}) reduces to
\begin{eqnarray}\label{cyclo_swsm}
\omega_{c_1}& =&\frac{B e v_F^2}{c \varepsilon _F}-\frac{s B^2 e^2 \hbar  v_F^4 \cos (\theta )}{2 c^2\varepsilon _F^3}
\end{eqnarray}
For a double-Weyl semimetal (\(J=2\)), we obtain
\begin{eqnarray}\label{cyclo_dwsm}
\omega_{c_2} &= &\frac{2 \alpha_2  B e \sin (\theta )}{c\hbar } + \frac{4 \alpha_2 ^2 B^2 e^2 s \cos ^3(\theta )}{c^2 \varepsilon_F  \hbar }
\end{eqnarray}
Finally, for a triple-Weyl semimetal (\(J=3\)), the cyclotron frequency becomes
\begin{eqnarray}\label{cyclo_twsm}
\omega_{c_3} &= & \frac{3 \alpha _3^{2/3} B e \varepsilon _F^{1/3} \sin^{\frac{4}{3}}(\theta )}{c \hbar ^{4/3}}\nonumber\\ &+& \frac{9 \alpha _3^{4/3} B^2 e^2 s \sin ^{\frac{2}{3}}(\theta ) (13 \cos (\theta )+3 \cos (3 \theta ))}{8 c^2
\hbar ^{5/3} \varepsilon _F^{1/3}}
\end{eqnarray}

Here, \(\varepsilon_F\) denotes the Fermi energy. Equations~(\ref{cyclo_swsm})--(\ref{cyclo_twsm}) demonstrate that the leading-order cyclotron frequency exhibits a nontrivial dependence on the topological charge. For the parameters considered here, the corresponding numerical estimates are
\(\omega_{c_1}^{(0)}
=1.8\times10^{11}~\mathrm{Hz}\),
\(\omega_{c_2}^{(0)}
=2.74\times10^{11}\sin\theta~\mathrm{Hz}\),
\(\omega_{c_3}^{(0)}
=1.14\times10^{11}\sin^{4/3}\theta~\mathrm{Hz}\). These results highlight the distinct angular dependence of the cyclotron motion in single- and multi-Weyl semimetals.\\
\onecolumngrid
\begin{center}\
\begin{figure}
\begin{tabular}{ccc}
\includegraphics[width=0.35\textwidth]{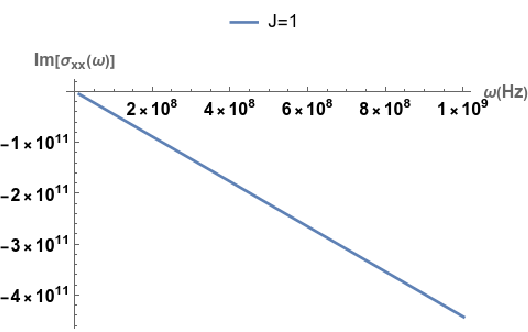} &
\includegraphics[width=0.35\textwidth]{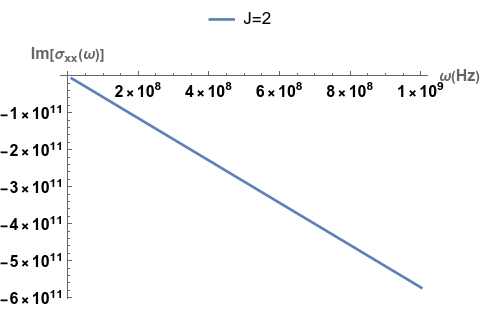} &
\includegraphics[width=0.35\textwidth]{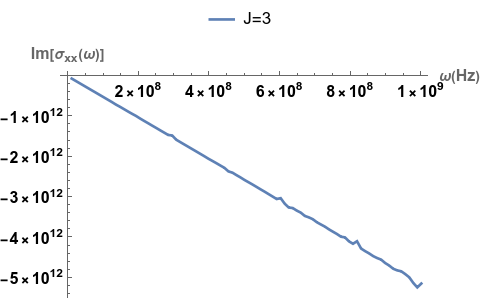} \\
\end{tabular}
\caption{The frequency dependence of the transverse optical conductivity at  B = 0.01 T. The other parameters are taken as same in Fig.(\ref{long_cond}).} 
\label{trans_cond}\
\end{figure}\
\end{center}\
\twocolumngrid

Having established the cyclotron frequencies, we proceed to evaluate the transverse optical conductivity. Unlike the longitudinal component, the transverse conductivity involves more complicated momentum-space integrals because of the angular dependence of both the group velocity and the cyclotron frequency. Consequently, compact analytical expressions are not readily obtainable, and we evaluate the corresponding integrals numerically.\\

\begin{widetext}
\begin{eqnarray}\label{cond_xx}
\sigma_{xx}^{(J)}(\omega)=i\frac{e}{8\pi^3}\int\limits_{-\infty}^{\infty}\, dk_z
\int\limits_0^{\infty}k_\perp dk_\perp \int\limits_{0}^{\pi}\, d\phi \cos \phi^2  \frac{e {v_{k_\perp}^s}^2}{2D}\Biggl(\frac{1}{\omega +\omega_{c_J}}+\frac{1}{\omega -\omega_{c_J}}\Biggr)\Biggl(-\frac{\partial f_0^s}{\partial \epsilon_{\bm k}^s}\Biggr) \\
\sigma_{xy}^{(J)}(\omega)=-\frac{e}{8\pi^3}\int\limits_{-\infty}^{\infty}\, dk_z
\int\limits_0^{\infty}k_\perp dk_\perp \int\limits_{0}^{\pi}\, d\phi \sin \phi^2 \frac{e {v_{k_\perp}^s}^2}{2D} \Biggl(\frac{1}{\omega + \omega_{c_J}}-\frac{1}{\omega-\omega_{c_J}}\Biggr)\Biggl(-\frac{\partial f_0^s}{\partial \epsilon_{\bm k}^s}\Biggr)
\end{eqnarray}
\end{widetext}

We evaluate these integrals numerically, retaining contributions up to quadratic order in the magnetic field. The resulting frequency dependence of the transverse and Hall conductivities is presented in Figs.~(\ref{trans_cond}) and (\ref{Hall_cond}), respectively. Both \(\sigma_{xx}^{(J)}\) and \(\sigma_{xy}^{(J)}\) exhibit an overall enhancement with increasing topological charge for the parameters considered.

Since our primary interest is the helicon propagation regime, we focus on the low-frequency limit, \(\omega\ll\omega_c\). In this regime, the cyclotron motion plays a dominant role in determining the transverse response. The remaining off-diagonal components of the conductivity tensor vanish owing to the symmetries of the system and the geometry considered here.\\

\section{Helicon dispersion relation}
We next obtain the dielectric tensor from Eq. (\ref{disp1.eqn}) using the constitutive relation $\mathbf D=\boldsymbol\epsilon\mathbf E$.

\begin{eqnarray}\label{dielectric}
\epsilon_{l m}= \delta_{l m}\epsilon_\infty + \frac{4\pi i}{\omega}\Bigl[\sigma_{l m}-\epsilon_{l m n}\frac{\alpha c}{2 \pi^2}\bigl(b_n-k_n\frac{b_0}{\omega}\bigr)\Bigr]
\end{eqnarray}

\noindent where $\epsilon_{l m n}$ is Levi-Civita antisymmetric tensor and the indices l,m and n run over the Cartesian coordinates x,y and z.  Combining this expression (\ref{dielectric}) with the electromagnetic wave equation (\ref{wave_equation}) gives

\begin{eqnarray}
\Bigl(\frac{c k}{\omega}\Bigr)^2-\frac{2\alpha}{\pi c \omega}\Bigl(b_z-\frac{b_0 k}{\omega}\Bigr)=\epsilon_\infty +\frac{4\pi i}{\omega} (\sigma_{xx}-i \sigma_{xy})
\end{eqnarray}

This expression describes the electromagnetic response resulting from the interplay between the conventional conductivity and the axion-electrodynamic contribution. In particular, the transverse longitudinal and Hall conductivity components modify the dielectric response, whereas the axion term introduces an additional dependence on the wave vector. Within the present approximation, the resulting dispersion relation contains both quadratic and linear contributions in the wave number. The corresponding dispersion curves for single-, double-, and triple-Weyl semimetals are presented in Fig.~(\ref{dis_plot}).\\

\onecolumngrid
\begin{center}\
\begin{figure}
\begin{tabular}{ccc}
\includegraphics[width=0.35\textwidth]{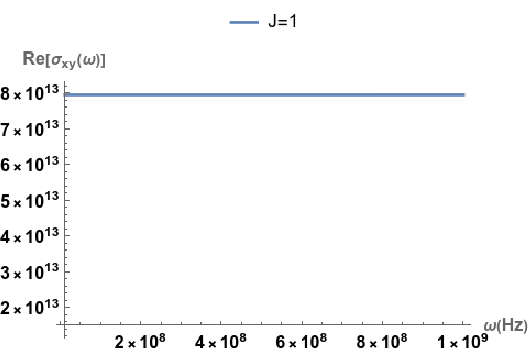} &
\includegraphics[width=0.35\textwidth]{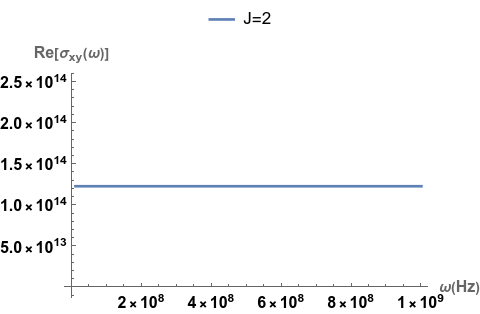} &
\includegraphics[width=0.35\textwidth]{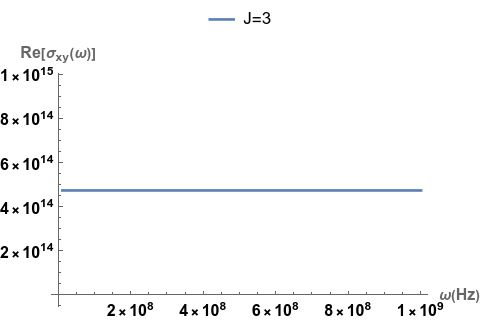} \\
\end{tabular}
\caption{The frequency dependence of the transverse Hall optical conductivity at  B = 0.01 T. The other parameters are taken as same in Fig.(\ref{long_cond}).} 
\label{Hall_cond}\
\end{figure}\
\end{center}\
\twocolumngrid

We now examine the gapped collective modes in the absence of an external magnetic field ($B=0$). In the long-wavelength limit, the collective excitation spectrum consists of three branches, labelled by $\lambda=1,2,3$, whose frequencies at zero wave vector are given by

\begin{eqnarray}
\Omega_{1,J}(k=0)=\omega_{-,J}\\
\Omega_{2,J}(k=0)=\omega_{p,J}/\sqrt{\epsilon_\infty}\\
\Omega_{3,J}(k=0)=\omega_{+,J}
\end{eqnarray}

\noindent where $\omega_{\pm,J}=\alpha c b/(\pi \epsilon_\infty) \pm \sqrt{
(\alpha c b)^2/(\pi \epsilon_\infty)^2+\omega_{p,J}^2/\epsilon_\infty}$ with $b=\vert\bm b\vert$
and $\omega_{p,J}$  denotes plasma frequency of a multi-Weyl semimetal with topological charge $J$. The dependence of the plasma frequency on the topological charge further highlights the role of the multi-Weyl band structure in determining the collective electromagnetic response. Using the zero-field longitudinal conductivity obtained in Eq.~(\ref{cond_xx}), the plasma frequency is expressed as

\begin{eqnarray}
\omega_{p,J}^2=\frac{4 J e^2\varepsilon_F^2}{3\pi\hbar^3v_F}
\end{eqnarray}

Thus, the axion contribution separates the two branches with frequencies $\omega_{-,J}$ and $\omega_{+,J}$, while the central branch is set by $\omega_{p,J}/\sqrt{\epsilon_\infty}$. The three branches should not be described as mutually degenerate when these frequencies differ. The numerical value quoted for $\omega_{p,1}$ should include its units and the parameter values used to obtain it.

\onecolumngrid\
\begin{center}\
\begin{figure}
\begin{tabular}{ccc}
\includegraphics[width=0.35\linewidth]{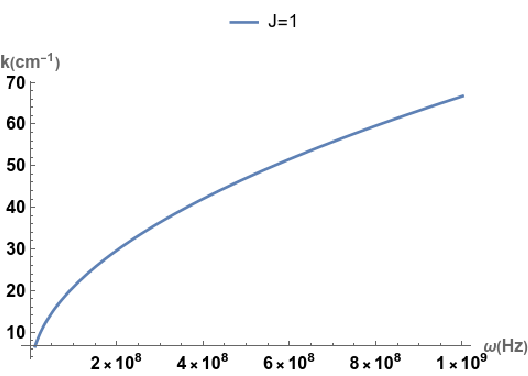} \
\includegraphics[width=0.35\linewidth]{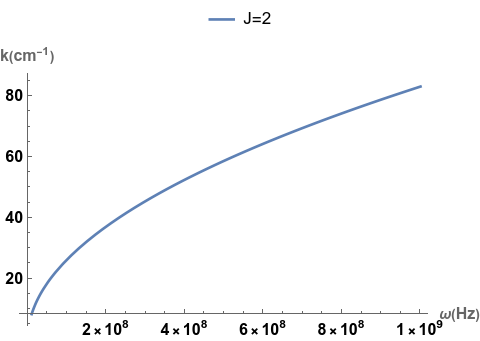} \
\includegraphics[width=0.35\linewidth]{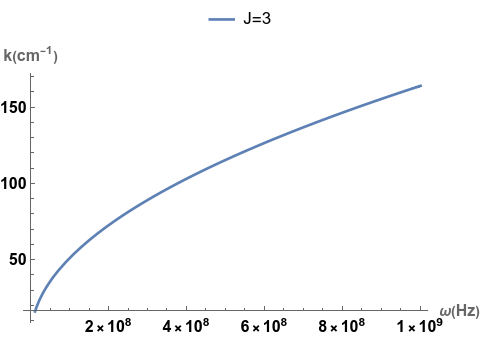}    
\end{tabular}
\caption{Dispersion relation $\omega$ versus $k$ for multi-Weyl semimetals at $B=0.01 \mathrm{T}$. The other parameters are $\epsilon_b=5$, $b_z=2\pi \times10^{10})\mathrm{cm}^{-1}$, $b_0=1.52 \times 10^{14}/sec $ and the fine-structure constant $\alpha=1/137$. All other parameters are the same as those used in Fig.~\ref{long_cond}.}\label{dis_plot}\
\end{figure}\
\end{center}\
\twocolumngrid\

\section{Conclusion}
We have investigated helicon excitations in three-dimensional multi-Weyl semimetals with $J=1,2,$ and $3$ using semiclassical Boltzmann transport theory, including OMM corrections to the quasiparticle dynamics. We obtained the longitudinal optical conductivity analytically and evaluated the transverse conductivity numerically in a weak static magnetic field. The resulting cyclotron dynamics depend on the topological charge and on the orbit geometry for the anisotropic bands. Within the dispersion approximation considered here, the helicon spectrum contains the conventional quadratic wave-number contribution and an additional term linear in wave number arising from axion electrodynamics. At zero magnetic field and in the long-wavelength limit, the collective spectrum has three gapped branches. The axion contribution separates the two axion-modified branches, while the mode frequencies also depend on the $J$-dependent plasma frequency. The results suggest that low-frequency electromagnetic propagation may provide a probe of the interplay between higher topological charge, anisotropic band structure, and topological electrodynamics. Quantitative comparison with experiment will require realistic material parameters and an assessment of scattering, damping, and the validity range of the semiclassical approximation.

\section*{Declarations}
\subsection{Funding}
 No external funding was received for this work.
\subsection{Data availability statement}
The figures were generated using Mathematica from the equations presented in the manuscript. The underlying data and Mathematica files are available from the corresponding author upon reasonable request.

\subsection{Competing interests}
The authors declare that they have no competing interests.\\

\subsection{Authors' contributions}
A.G. conceived the study and formulated the theoretical framework. S.K Ram performed the analytical calculations and numerical analysis. A.G. and S.K Ram discussed the results and contributed to the interpretation of the findings. Both authors contributed to the writing and revision of the manuscript and approved the final version.\\

\section{Acknowledgements}
We thank Debanand Sa for useful discussions. 

\bibliography{multi_helicon}
\end{document}